\documentclass{article}
\usepackage{spconf,amsmath,graphicx,booktabs}

\title{Multimodal Emotion Recognition from Raw Audio with Sinc-convolution}
\twoauthors
{Xiaohui Zhang, Mangui Liang}
{Department of Computer Science \\
	Beijing JiaoTong University \\
	Beijing 100044, China}
{Wenjie Fu}
{Research Center of 6G Mobile Communications, \\
	School of Cyber Science and Engineering, \\ 
	and Wuhan National Laboratory for Optoelectronics \\
	Huazhong University of Science and Technology \\
	Wuhan 430074, China}
\begin{document}
	%
	\maketitle
	\begin{abstract}
		Speech Emotion Recognition (SER) is still a complex task for computers with average recall rates usually about 70\% on the most realistic datasets. Most SER systems use hand-crafted features extracted from audio signal such as energy, zero crossing rate, spectral information, prosodic, mel frequency cepstral coefficient (MFCC), and so on. More recently, using raw waveform for training neural network is becoming an emerging trend. This approach is advantageous as it eliminates the feature extraction pipeline. Learning from time-domain signal has shown good results for tasks such as speech recognition, speaker verification etc. In this paper, we utilize Sinc-convolution layer, which is an efficient architecture for preprocessing raw speech waveform for emotion recognition, to extract acoustic features from raw audio signals followed by a long short-term memory (LSTM). We also incorporate linguistic features and append a dialogical emotion decoding (DED) strategy. Our approach achieves a weighted accuracy of 85.1\% in four class emotion on the Interactive Emotional Dyadic Motion Capture (IEMOCAP) dataset.
	\end{abstract}
	\section{INTRODUCTION}
	\label{sec:intro}
	
	Speech emotion recognition (SER) is a major research area. Most of the acoustic features used in the fusing system are hand-crafted features \cite{Fac15}, such as MFCC, pitch and voice quality features extracted from the raw audio or features learned automatically from spectrograms \cite{Pepino20-DAF}, \cite{Satt17-EER}. However, the hand-crafted features contain more limited information than the raw audio signal, which makes more and more researchers pay attention to extracting features from raw audio more flexibly. For example, Guizzo et el \cite{Guizzo20-MTS} present a multi-time-scale (MTS) convolution layer which does not increase the number of parameters but increases the temporal flexibility compared to standard CNNs. Xu et el \cite{Hgfm} propose a hierarchical grained and feature model (HGFM) which includes a frame-level representation module with before and after information, a utterance-level representation module with context information, and a different level acoustic feature fusing module. In this paper, we append the sentences of raw audio as acoustic features to our model. 
	\par
	Due to the great success of CNN in the field of speech emotion recognition, most systems use the convolution layer to process acoustic features. However, the standard CNN convolution layer has many filter parameters, which makes the convolution layer only capable of learning the low dimensional and compacted features \cite{zhang2023remember, zhang23DoYou}. To help the input layer discover more meaningful filters, Ravanelli el et \cite{Ravanelli18-SRF} proposes SincNet, a neural architecture of directly processing waveform audio. The Sinc-convolution layer of SincNet is not only faster in convergence speed than a standard CNN but also more computationally efficient due to the exploitation of filter symmetry. For this reason, we utilize the Sinc-convolution layer as the input layer to extract acoustic information from raw audio signals, and we utilize LSTM followed by the Sinc-convolution layer to extract information between sentences. Compared with conventional convolution layers, Sinc-convolution layers can learn the parameters of filters from raw audio signals, have fast convergence and higher interpretability with a smaller number of parameters. Humans express emotions through multi-modal ways, which suggests that we can process multi-modal features by fusing model to improve the accuracy of predictions \cite{zhang2023multimodal}. At present, the feature-level fusion of acoustic features and linguistic features is more commonly used. Pepino et el \cite{Pepino20-DAF} present different fusing models for SER by combining acoustic and linguistic features. The result shows that multi-modal system leads to significant improvements of approximately $16\%$ on Interactive Emotional Dyadic Motion Capture (IEMOCAP) \cite{Busso08-IEM}. In this paper, we study a multimodal emotion recognition model with combined acoustic and linguistic information and utilize dialogical emotion decoder (DED) \cite{Yeh20-DED} to process the prediction result of pre-trained classifier. DED decodes each utterance into one of the four emotion categories at inference stage. Experiments are performed on IEMOCAP dataset. Our approach achieves a weighted accuracy of $85.1\%$ in four class emotion on IEMOCAP.
	
	\section{METHODOLOGY}
	\label{sec:Method}
	\subsection{Acoustic Model}
	\label{ssec:acoustic}
	
\begin{figure}[htb]
		\begin{minipage}[b]{1.0\linewidth}
			\centering
			\centerline{\includegraphics[width=8.5cm]{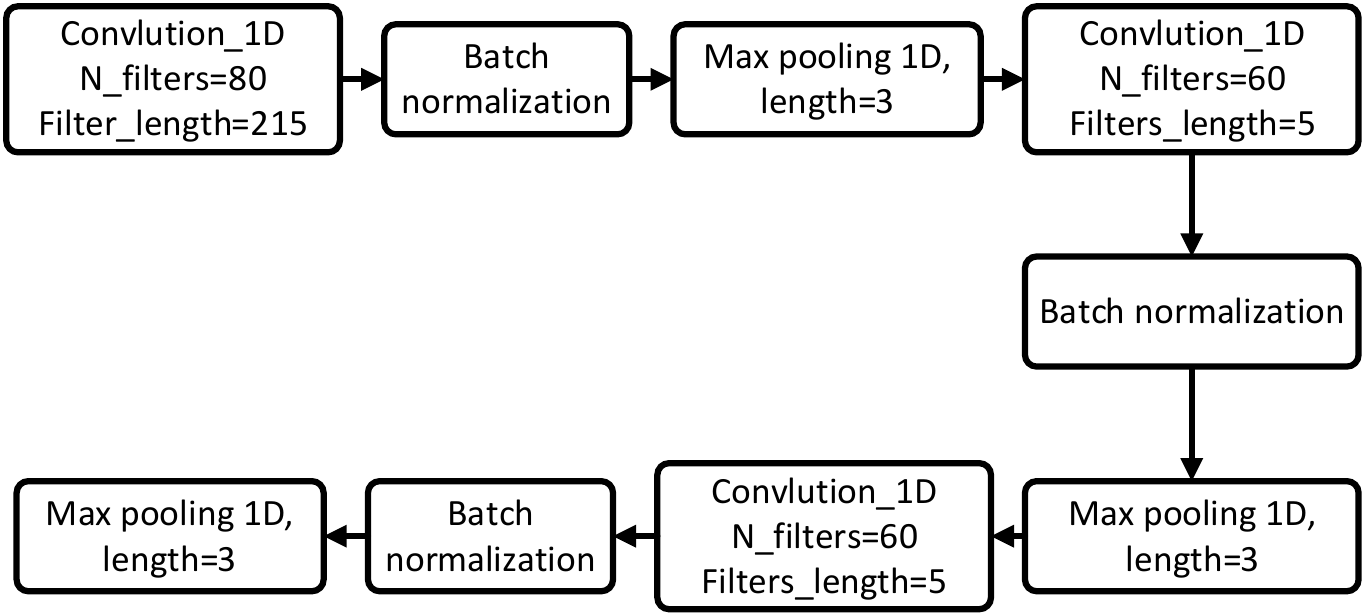}}
		\end{minipage}
		\caption{Architecture of Sinc-conv layer.}
		\label{fig:Sinc_layer}
	\end{figure}

	To learn more useful information from raw signals, we utilize Sinc-convolution filter layers \cite{Ravanelli18-SRF} to learn custom filter banks tuned for emotion recognition from speech audio. Neural architecture for processing raw audio samples as shown in Figure \ref{fig:Sinc_layer}.
	
	The first layer of a standard CNN performs a set of time-domain convolutions between the input waveform and some Finite Impulse Response (FIR) filters. Each convolution is defined as follows: 
	\begin{equation}
	y[n] = x[n]*h[n] = \sum\limits_{l = 0}^{L - 1} x[l]\cdot h[n - l]
	\label{eq1}
	\end{equation}
	where $x[n]$ is a chunk of speech signal. $h[n]$ is the filter of length $L$, and $y[n]$ is the filtered output. In conventional convolution layer, all the $L$ elements of each filter are learned from data. Conversely, the proposed Sinc-convolution layer performs the convolution with a pre-defined function $g$ that depends on fewer learnable parameters $\theta$ only \cite{Ravanelli18-SRF}
	\begin{equation}
	y[n] = x[n]*g[n,\theta]
	\label{eq2}
	\end{equation}
	Each defined filter-bank is composed of rectangular band-pass filters. In the frequency domain, the magnitude of a generic band-pass filter can be represented by two low-pass filters. 
	\begin{equation}
	G[f,f_1 ,f_2] = {\rm rect}\left(\frac{f}{2f_2} \right) - {\rm rect}\left(\frac{f}{2f_1} \right)
	\label{eq3}
	\end{equation}
	where $f_1,f_2$ refers to low and high cutoff frequencies and $\rm rect$ function is the rectangular function in the magnitude frequency domain. The time-domain representation of the function $g$ can be derived as follows
	\begin{equation}
	g[n,f_1 ,f_2] = 2f_2 {\rm sinc}(2\pi f_2 n) - 2f_1 {\rm sinc}(2\pi f_1 n)
	\label{eq4}
	\end{equation}
	where the $\rm sinc$ function is defined as ${\rm sinc}(x)=\sin(x)/x$.
	We compare Sinc-convolution layers followed by Deep Neural Network (DNN) and LSTM. We also compare conventional convolution layers of CNN fed by spectrogram computed from raw audio and Sinc-convolution layers of Sinc-DNN and Sinc-LSTM fed by 250ms chunk selected from raw audio randomly. Though CNN is based on the same architecture as Sinc-DNN, it replaces the sinc-based convolution with a standard one. Both Sinc-convolution layers and convolution layers of CNN use Batch Normalization. The performance is evaluated with Sentence Error Rate (SER), which represents the average of error rate of each sentence.
	
\begin{figure}[htb]
		\begin{minipage}[b]{1.0\linewidth}
			\centering
			\centerline{\includegraphics[width=8.5cm]{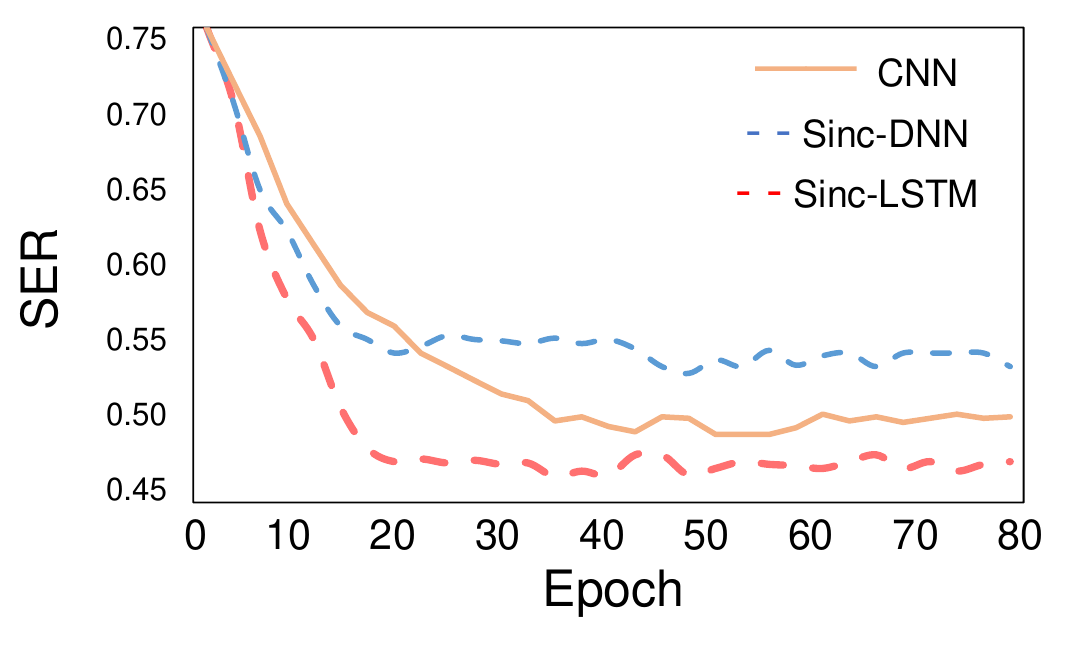}}
		\end{minipage}
		\caption{Sentence Error Rate of CNN, Sinc-DNN and Sinc-LSTM over various training epochs}
		\label{fig:Cm_Sinc}
	\end{figure}

	The learning curves of Sinc-DNN and CNN can be observed from Figure \ref{fig:Cm_Sinc}. Both convolution layer of Sinc-DNN and standard CNN followed by a Deep Neural Network (DNN), which contains several fully connected layers and Batch Normalization. Figure \ref{fig:Cm_Sinc} shows that both Sinc-DNN and Sinc-LSTM have faster convergence than CNN beacuse they reduces the number of parameters in the first convolutional layer and the function $g$ is symmetric that means we only need to consider one side of the filter and inherit the result for the other half.
	
	However, the CNN converges to a better performance leading to a Sentence Error Rate (SER) of 46\% against a SER of 51\% achieved with the Sinc-DNN. One reason for this result is that Sinc-DNN does not pay attention to the information between sentences, which is helpful in the emotion recognition. We learn this neglected information by replacing DNN with LSTM. Another reason is that some of the extracted features may be meaningless noise due to our random chunk extractions from raw audio. To reduce the noise interference caused by randomly extracting features, we extract the features after pre-processing the signals in the database, such as extracting the parts with higher power to effectively avoid the noise interference. As observed in the Figure \ref{fig:Cm_Sinc}, the SER of Sinc-LSTM dropped to 43\%, which is 8\% lower than Sinc-DNN and 3\% lower than CNN.
	
	\subsection{Linguistic Model}
	\label{ssec:linguistic}
	Recurrent Neural Network (RNN) is widely used in natural language processing tasks. However, RNNs cannot remember longer sentences and sequences due to the vanishing/exploding gradient problem. It can only remember the parts which it has just seen. The emergence of LSTM has solved this problem well \cite{Huang17-CVL}. Although an LSTM is supposed to capture the long-range dependency better than the RNN, it is difficult to give large weights to the important words for estimation. To this end, we deploy a self-attention layer on top of the LSTM. The self-attention mechanism can focus on the important parts in the sentence \cite{Xie19-SAB}. The more relevant the semantic relation between words and emotion is, the greater the weight of the connection between them, which can assist the model to determine more accurate classification. 
	\subsection{Feature Fusion}
	\label{ssec:modelfusion}
	Feature-level fusion is used to fuse linguistic and acoustic features obtained from individual networks. A 2048-D feature vector from the acoustic network and a 4800-D feature vector from the linguistic network are concatenated. We use attention before fusion where attention is applied on individual feature vectors. Finally, the utterance emotion is classified with the use of a “Softmax” activation and a dialogical emotion decoder (DED) which is a post-processing over the final dense layer of the fusing network.
	
	In this paper, we study two fusing models without DED. One is the acoustic feature of Sinc-DNN fusing with the linguistic feature of LSTM (M1), and the other is the acoustic feature of Sinc-LSTM fusing with the linguistic feature of LSTM (M2). We also utilize DED to classify the prediction of M2, which makes the performance better than M2. 
	\subsection{Dialogical Emotion Decoder}
	The decoding process of Dialogical Emotion Decoder (DED) is built on top of a SER model which is an approximate inference algorithm which decodes each utterance into one of the four emotion categories at inference stage. This decoder is built on three core ideas: the emotion which occurs more frequently in dialog history is more likely to show up again; while not all utterances have consensus labels; and the posterior distributions capturing affective information would enable us to decode utterances in sequence. DED can model the emotion flows in a dialog consecutively by combining emotion classification, emotion shift and emotion assignment process together. Yeh et el \cite{Yeh20-DED} achieves $70.1\%$ unweighted accuracy on four emotion class in the IEMOCAP.
	
	\section{EXPERIMENTS}
	\label{sec:experiment}
	\subsection{Experiment Setup}
	\label{ssec:setup}
	
	In this work, we used the IEMOCAP, a benchmark dataset containing about 12 hours of audio and video data, as well as text transcriptions. The dataset contains five sessions, each of which involves two distinct professional actors conversing with one another in both scripted and improvised manners. In this work, we utilize data from both scripted and improvised conversations, as well as mere audio data to stay consistent with the vast majority of prior work. We also train and evaluate our model on four emotions: happy, neutral, angry, and sad, resulting in a total of 5531 utterances (happy: 29.5\%, neutral: 30.8\%, angry: 19.9\%, sad: 19.5\%).
	
	\subsection{Result and Analysis}
	\label{ssec:res}
	
	We compare the performance of three models: 1) a standard CNN; 2) a Sinc-DNN; and 3) a Sinc-LSTM.
	\begin{table}[th]
		\caption{Performance of CNN, Sinc-DNN and Sinc-LSTM.}
		\label{tab1}
		\centering
		\setlength{\tabcolsep}{1.1mm}
		\begin{tabular}{ cccc@{}l  r }
			\toprule
			\multicolumn{1}{c}{\textbf{Model}} &
			\multicolumn{1}{c}{\textbf{Training time($epoch$)}} &
			\multicolumn{1}{c}{\textbf{WA($\%$)}} & 
			\multicolumn{1}{c}{\textbf{UA($\%$)}} \\
			\midrule
			CNN              &$40$                & $54.3$  & $53.2$~~~       \\
			Sinc-DNN      &$25$                & $49.5$  & $47.8$~~~       \\
			Sinc-LSTM     &$25$                & $57.1$  & $54.5$~~~       \\
			\bottomrule
		\end{tabular}
	\end{table}
	It can be seen from Table \ref{tab1} that Sinc-LSTM outperforms CNN with a $2.8\%$ improvement that is also obtained with faster convergence (25 vs 40 epochs). Sinc-convolution layer is specifically designed to implement rectangular bandpass filters, leading to more meaningful CNN filters, and LSTM can learn more contextual information than DNN does. This makes Sinc-LSTM perform better.
	The performance of multi-modal systems is indicated in Table \ref{tab2}. CNN+LSTM is the our baseline with acoustic model CNN and linguistic model LSTM. MDRE \cite{Yoon18-MSE} using dual RNNs encode acoustic and linguistic information and predict using a feed-forward neural model. MHA-2 \cite{Yoon19-SER} propose a multi-hop attention model which is designed to process the contextual information from transcripts and audio. Table \ref{tab2} shows that MHA-2 has outperformed CNN+LSTM by $5.7\%$ in the terms of WA when the model is applied to the IEMOCAP dataset and MHA-2 has outperformed MDRE by $4.7\%$ that demonstrates the attention framework can increase performance. 
	\begin{table}[th]
		\caption{Performance of different models.}
		\label{tab2}
		\setlength{\tabcolsep}{1.0mm}
		\begin{tabular}{ cccc@{}l  r }
			\toprule
			\multicolumn{1}{c}{\textbf{Model}} &
			\multicolumn{1}{c}{\textbf{Modality}} &
			\multicolumn{1}{c}{\textbf{WA($\%$)}} & 
			\multicolumn{1}{c}{\textbf{UA($\%$)}} \\
			\midrule
			CNN+LSTM                              &Audio+Text       & $70.8$  & $69.7$~~~       \\
			MDRE \cite{Yoon18-MSE}				  &Audio+Text		 & $71.8$  & $-$~~~          \\
			MHA-2 \cite{Yoon19-SER}      		  &Audio+Text       & $76.5$  & $77.6$~~~       \\	
			M1=Sinc-DNN+LSTM     			  &Audio+Text       & $72.1$  & $71.1$~~~       \\
			M2=Sinc-LSTM+LSTM     			  &Audio+Text       & $75.9$  & $76.3$~~~       \\
			IAAN \cite{Yeh19-IAA}     			  &Audio         & $65.2$  & $67.1$~~~       \\
			IANN+DED \cite{Yeh20-DED}     		  &Audio         & $69.0$  & $70.1$~~~       \\
			M2+DED     							  &Audio+Text       & $85.3$  & $85.1$~~~       \\
			\bottomrule
		\end{tabular}
		
	\end{table}
	
	\begin{figure}[htb]
		\begin{minipage}[b]{1.0\linewidth}
			\centering
			\centerline{\includegraphics[width=8.5cm]{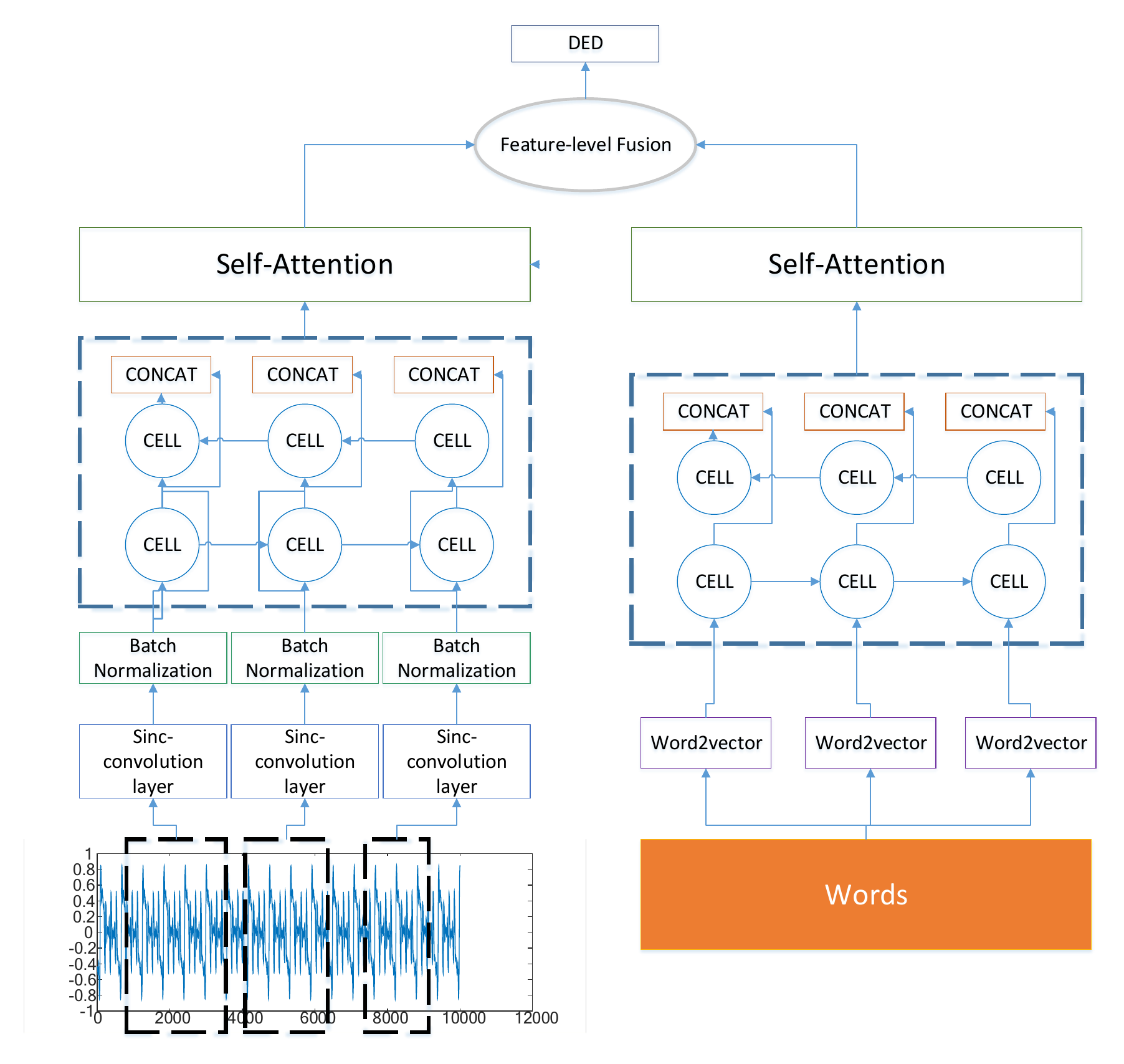}}
		\end{minipage}
		\caption{The Architecture of fusing model combining Sinc-LSTM and LSTM with DED post-processing}
		\label{fig:img3}
	\end{figure}
	
\begin{figure}[htb]
		\begin{minipage}[b]{1.0\linewidth}
			\centering
			\centerline{\includegraphics[width=8.5cm]{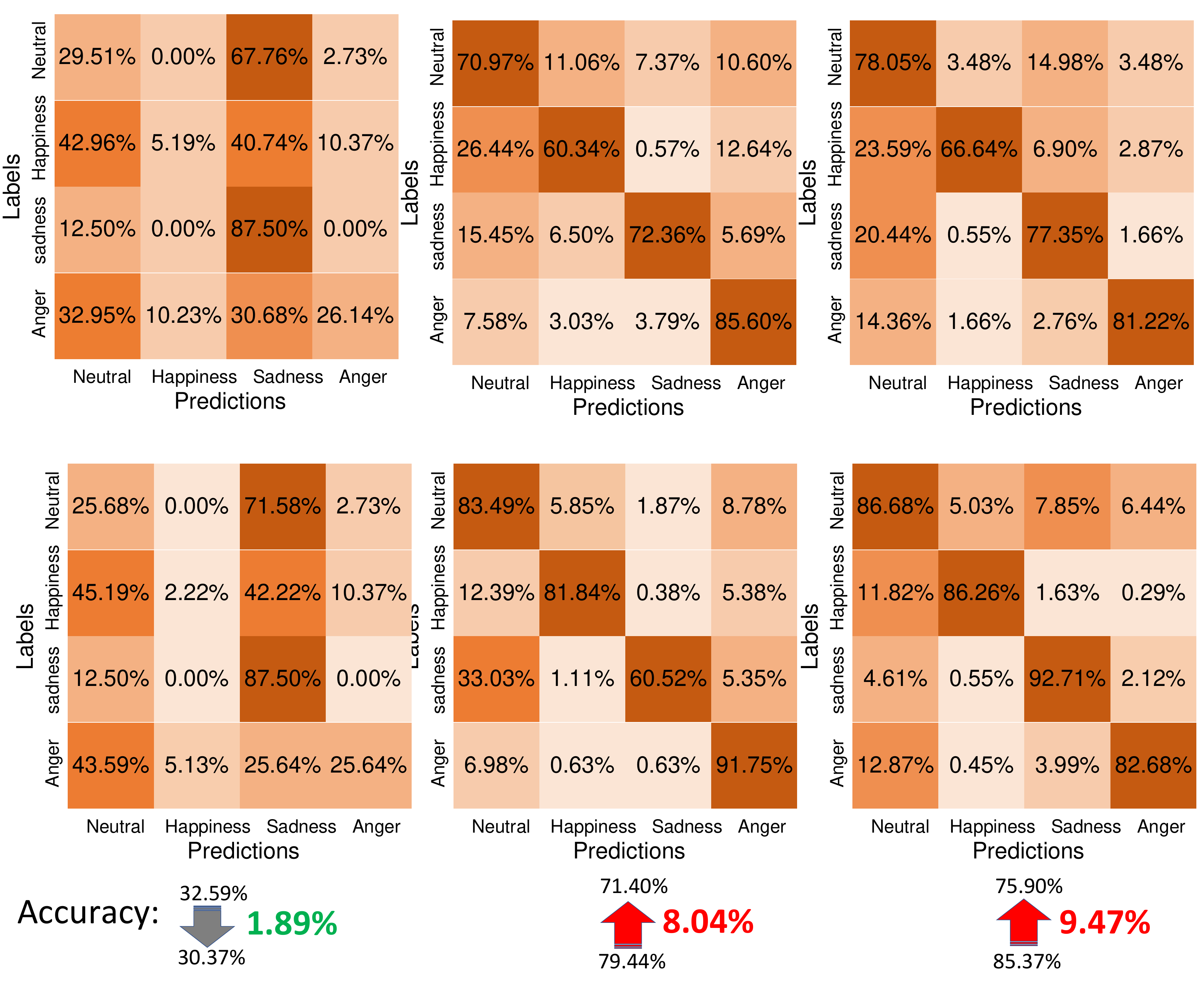}}
		\end{minipage}
		\caption{The performance of DED with different pre-classifiers}
		\label{fig:ded}
	\end{figure}

	As observed from Table \ref{tab2}, Both M1 and M2 perform better than CNN+LSTM, which indicates that the multi-modal model using Sinc-convolution layer not only has faster convergence than that using CNN, but also has better verification accuracy. The unweighted accuracy of Sinc-DNN is lower than CNN but M1 has outperformed CNN+LSTM by 1.3\%. The main reason is that the information between sentences can learned from textual features by LSTM. The M2 system using LSTM performs better than M1. Specifically, M2 achieves 75\% accuracy, with relative 3\% improvement over M1, which verifies that LSTM is more suitable for processing contextual information on IEMOCAP than DNN. Figure \ref{fig:img3} shows the architecture of M2+DED. Based on the M2 system, we use DED to classify the prediction of M2, and finally achieves an accuracy rate of 85\%, which improves the accuracy rate by 8.8\% compared with the MHA-2. However, as shown in Table \ref{tab2}, the effectiveness of IAAN+DED is not as significant as M2+DED. First, DED relies on a well-performing classifier \cite{Yeh20-DED}. Due to the poorer performance of IAAN in the IEMOCAP, the effect of DED is more limited. Furthermore, as the IAAN+DED only extract acoustic low-level descriptors (LLD), including features such as MFCCs, the ability of DED in properly decoding through long sequence of dialogs is not as obvious as for the raw audio. From Figure \ref{fig:ded}, We also find that the performance of DED is more significant when accuracy of the pre-classifier is higher than $30\%$ because DED classifies the current emotion based on the previous and the current prediction of the pre-classifier. If the correct account for a large proportion of all predictions reaches, for example, $75\%$ of M2, then even if the current is wrong, it can still be corrected by the previous correct predictions.
	
	\section{CONCLUSIONS}
	\label{sec:conclusion}
	
	In this paper, we utilize Sinc-convolution layer to extract acoustic features from raw audio followed by an LSTM and a multi-modal emotion recognition system based on Sinc-LSTM and LSTM combining acoustic and linguistic data (Sinc-LSTM+LSTM). The performance of Sinc-LSTM+LSTM is better than CNN+LSTM, which shows that not only acoustic model but also fusing model using Sinc-convolution layer has higher verification accuracy. We do not use the hand-crafted features that have been used in the standard CNN but use the raw audio append to the Sinc-convolution layer, which makes Sinc-LSTM+LSTM learn more information than the baseline using CNN and LSTM does. Finally, the prediction of Sinc-LSTM+LSTM is appended to dialogical emotion decoding (DED) strategy, a post-processing that classifies the current result based on the previous and current prediction of pre-classifier. The performance of DED has achieved a weighted accuracy of 85.29\% on the IEMOCAP database, which outperforms the Sinc-LSTM+LSTM by 9.96\%.
	
	
	
	\vfill\pagebreak
	
	\bibliographystyle{IEEEbib}
	\bibliography{zhang}
	
\end{document}